\documentclass[twocolumn]{aastex701}

\usepackage{hyperref}
\usepackage{natbib}
\usepackage{lipsum}
\usepackage{amsmath}

\newcommand{\msun}{\ensuremath{\,M_\odot}}

\newcommand{\teff}{{$T_{\rm eff}$}}
\newcommand{\mstar}{${M}_{\star}$}

\newcommand{\logg}{$\log{g}$}
\newcommand{\gmag}{$M_G$}
\newcommand{\deltaP}{$\Delta\Pi^a_1$}

\newcommand{\targ}{WD\,0158$-$160}

\newcommand{\cuny}{Department of Physics, City University of New York Graduate Center, 365~5th~Ave, New~York, NY~10016, USA}
\newcommand{\queens}{Physics Department of Queens College, Queens College Science Building, 6530~Kissena~Blvd~B334, Queens, NY~11367, USA}

\graphicspath{{./}{figures/}}

\submitjournal{AAS Journals}

\shorttitle{The seismic technique for white dwarf fundamental parameters}
\shortauthors{Bell \& Bischoff-Kim}

\begin{document}

\title{A Seismic Technique for Obtaining White Dwarf Fundamental Parameters\\ from Mean Period Spacings and Astrometry}

\author[0000-0002-0656-032X]{Keaton J.\ Bell}
\affiliation{\queens}
\affiliation{\cuny}
\affiliation{Department of Astrophysics, American Museum of Natural History, Central Park West at 79th Street, New York, NY 10024, USA}
\email{keaton.bell@qc.cuny.edu}

\author[0000-0002-7487-9340]{Agn{\`e}s Bischoff-Kim}
\affil{Penn State Wilkes-Barre,
44 University Drive,
Dallas, PA 18612, USA}
\email{xk55@psu.edu}

\begin{abstract}

We present a new statistical technique that utilizes the synergy of precision astrometry and time series photometry from modern space missions to obtain reliable physical parameters of pulsating white dwarf stars. We compute a grid of white dwarf structural models that span the helium-atmosphere pulsating white dwarf (DBV) instability strip, showing that mean period spacings between adjacent pulsation modes and absolute magnitudes derived from \textit{Gaia} astrometry vary monotonically and in opposing directions across parameter space. While most efforts in white dwarf asteroseismology to directly fit individual pulsation periods to stellar models result in degenerate and poorly resolved solutions, the ``seismic technique'' produces unique and reliable seismic solutions for global parameters of mass and effective temperature when a reliable mean period spacing is detected. Our models sample various physically plausible interior chemical composition profiles based on modern evolutionary models to propagate uncertainty from the precise structures of actual stars. Once the global stellar parameters are tightly constrained, seismically resolving white dwarf interior structures becomes more computationally tractable, and degeneracies can be resolved. This new seismic technique is largely insensitive to the precise absorption line profiles interpreted by the widely used spectroscopic technique, and therefore provides complementary constraints that can be used to test spectroscopic methods. We demonstrate the method for the pulsating helium-atmosphere white dwarf \targ\ observed by TESS, obtaining $T_\mathrm{eff} = 24584\pm 971$\,K and $M_\star = 0.608\pm 0.013$\,\msun.

\end{abstract}

\keywords{\uat{Asteroseismology}{73} --- \uat{Astrometry}{80} --- \uat{Fundamental parameters of stars}{555} --- \uat{White dwarf stars}{1799}}

\section{Introduction} \label{sec:intro}

One of the most widely relied upon methods for determining the physical parameters of white dwarf stars is the ``spectroscopic technique,'' by which measured absorption line profiles are fit against a grid of simulated model atmospheres to determine effective temperatures and surface gravities.  These can be compared with structural and evolutionary models to infer white dwarf masses, radii, and cooling ages.
This method was first employed to determine the atmospheric properties of a handful of pulsating hydrogen-atmosphere white dwarf stars by \citet{Bergeron1990} and \citet{Daou1990}, and later extended by \citet{Bergeron1992} to measure the mass distribution of hydrogen-atmosphere white dwarfs more broadly. The spectroscopic technique has since become the preferred method for obtaining precision parameters for all types of white dwarfs, with wholesale application to, e.g., Sloan Digital Sky Survey spectra \citep[$>30{,}000$ white dwarfs spectroscopically characterized;][]{Kepler2019}.

Even for the simplest case of white dwarfs with pure-hydrogen atmospheres, there are several reasons to be skeptical of the accuracy of parameters derived with the spectroscopic method.  For one, different treatments of the atomic physics yield inconsistent results, prompting continual improvement of the physical models. Recent advancements include the improved treatment of Stark broadening \citep{Tremblay2009,Gomez2017} and 3D simulations of convection \citep{Tremblay2013,Cukanovaite2018}, though systematic offsets between spectroscopically derived parameters and values obtained by other means persist (e.g., gravitational redshift, \citealt{Falcon2010}; the multi-color photometric technique, \citealt{Genest-Beaulieu2019,Tremblay2019}). Fitting modern theoretical line profiles to absorption spectra measured at white dwarf photospheric conditions in laboratory experiments also yields inconsistent electron densities \citep{Schaeuble2019}.
In addition, the details of how the white dwarf spectra are obtained, reduced, and compared to models can significantly influence the inferred parameters.  \citet[Section 4.5]{Fuchs2017} quantified ten such sources of systematics for white dwarf spectra obtained with the same instrumental setup, including how they are flat fielded, where around the absorption lines the spectra are continuum normalized, and which spectral lines are included in the fits.

There are numerous other methods for constraining white dwarf parameters that are often used in combination with the spectroscopic method. Among these are astrometry
and asteroseismology. 
In this work, we connect these two approaches directly to develop a ``seismic technique''\footnote{Referred to by analogy to the spectroscopic technique \citep{Bergeron1990} and photometric technique \citep{Bergeron1997}, which are specific statistical approaches to interpreting spectroscopic and photometric data to constrain global white dwarf parameters of effective temperature and mass (or surface gravity).}
for deriving white dwarf parameters that is largely independent of the spectroscopic technique. 
This statistical seismic technique is expected to produce more accurate measurements of white dwarf mass and effective temperature, due to the relative simplicity of interpreting the relevant observables (compared to spectral line profiles), and with reported errors propagated from uncertainty on the precise white dwarf interior structure within expectations from evolutionary models \citep[e.g.,][]{Althaus2010,DeGeronimo2017}.
The results can therefore be used to calibrate model atmosphere physics and fitting methods to improve the accuracy of the spectroscopic technique, which is applicable to a much larger sample of (mostly non-pulsating) white dwarfs.

Distance constraints from precise astrometric parallaxes measured by \textit{Gaia} \citep{GaiaEDR3} enable the conversion from apparent to absolute magnitudes. 
A measured absolute magnitude constrains radius as a function of effective temperature, since a smaller, hotter star could have the same magnitude as a larger, cooler star. White dwarfs follow a mass-radius relationship where more massive white dwarfs are smaller, so astrometric constraints present as a trend line in the mass-effective temperature plane. One recourse to further determine the mass is to constrain temperatures with multi-band photometric colors \citep[the photometric technique; e.g.,][]{Bergeron1997,Hollands2018,Genest-Beaulieu2019,Bergeron2019}. Our seismic technique instead relies on observed pulsations to break this degeneracy.

The properties of pulsating white dwarf stars can be analyzed using the tools of asteroseismology \citep{Winget2008,Corsico2019,Bell2026}. 
Stellar pulsations oscillate as standing waves within a star at eigenfrequencies tuned by the precise stellar interior. These pulsations cause brightness variations at stellar resonant frequencies, which can be measured from time series photometry. In principle, the precise radial chemical profile of a star can be constrained by fitting individual measured pulsation periods to values calculated for either evolutionary \citep[e.g.,][]{Romero2017} or parameterized \citep[e.g.,][]{Giammichele2022} structure models.
In practice, the standard period-by-period fitting approach often admits degenerate solutions \citep[due to uncertainty on which pulsation modes are observed, or from effects like core/envelope symmetry;][]{Montgomery2003}, and some pulsating white dwarfs exhibit fewer modes than the number of independent free parameters in the models. In many asteroseismic analyses of pulsating white dwarfs, the solution space is not sufficiently resolved to yield reliable mass and temperature determinations (\citealt{Charpinet2015}; A.~Dublin et al., submitted).
These issues do not affect the comparison of the mean spacing between pulsation periods of gravity modes with consecutive radial orders to the asymptotic spacing calculated from models, which for helium-atmosphere white dwarfs is most sensitive to effective temperature and stellar mass \citep{Tassoul1990}.  For a white dwarf, the mean period spacing typically increases as the star cools. More massive white dwarfs exhibit pulsation spectra with smaller period spacings. 

The astrometric and mean period spacing constraints cross the mass versus effective temperature plane diagonally in opposing directions, intersecting at a unique location. Combined, they enable us to establish global parameters of pulsating white dwarfs that are largely independent of the spectroscopic technique. This seismic technique provides another angle for characterizing the global parameters of individual pulsating white dwarfs that exhibit a clear sequence of radial overtones.  We establish this methodology in this work, demonstrating how we determine the mass and effective temperature of the pulsating helium-atmosphere white dwarf star WD\,0158$-$160 that has had its pulsation frequencies precisely measured \citep{Bell2019} from data from the Transiting Exoplanet Survey Satellite \citep[\emph{TESS};][]{Ricker2014}. The methodology attempts to achieve realistic uncertainties by propagating errors from multiple potential sources, including uncertainty on the precise interior structure of the star evaluated by computing models that span the range of composition profiles produced by fully evolutionary models in the literature. Employing this technique on a larger sample of pulsating white dwarfs will help to calibrate the spectroscopic method, as well as enable more accurate asteroseismic constraints of white dwarf internal structures.

\section{The data}\label{sec:target}

Our seismic technique for the statistical determination of masses and effective temperatures for pulsating white dwarfs relies on the interpretation of two key observables: measured mean period spacings within a radial overtone series of gravity-mode pulsations (\deltaP) and absolute magnitudes established with astrometry (\gmag). This paper demonstrates the methodology for analyzing the bright helium-atmosphere pulsating white dwarf (DBV) star \targ\ \citep{Kilkenny2016}, for which suitable measurements are available. This $V = 14.5$\,mag star is also known as G\,272-B2A and EC 01585-1600 and is located at Right Ascension 30.23689\,deg, Declination $-15.76923$\,deg. 

\targ\ was the subject of a TESS first-light paper from Working Group 8 (compact pulsators) of the TESS Asteroseismic Research Consortium\footnote{\url{https://tasoc.dk/}} \citep{Bell2019}. It is assigned the identifier TIC\,257459955 in the TESS Input Catalog \citep{Stassun2019}. This target was featured for its particularly rich set of nine independent pulsation modes detected in the 2-minute cadence TESS light curve from Sector 3. Many of these pulsation signals appear to align with a regular spacing in period. This is the signature of a radial overtone sequence (labeled\footnote{Often labeled as $n$, especially for other types of pulsating star.} as radial order $k$) for gravity modes of the same spherical degree ($\ell$) in the asymptotic limit \citep[$k\gg\ell$; for an overview of white dwarf pulsation characteristics see][]{Bell2026}. Multiple statistical tests applied to detected pulsation periods longer than 485\,s \citep[$k\gtrsim10$, chosen to avoid strongly trapped modes;][]{Corsico2002} support that these signals belong to a sequence with a mean period spacing of 38.1\,s \citep{Bell2019}. This is consistent with the expected spacings of $\ell=1$ modes, and these are also the modes that are expected to be most visible in disc-integrated light due to having low geometric cancellation \citep{Dziembowski1977}. Modes of given $\ell$ and $k$ can be split into a maximum of $2\ell+1$ frequency multiplets by rotation (each with different azimuthal order, $m$); however, in the TESS data on \targ\ we only observe one arbitrary $m$ component for all but one of these signals. Because the mean period spacing is most robustly determined from the central ($m=0$) components of the multiplets, the specific components observed could affect the slope of the best-fit line to the periods versus relative overtone number. To propagate the error from the unknown $m$ identifications of the detected modes in \targ, \citet{Bell2019} randomly sample from potential $m$ values for each mode to determine a measured mean period spacing with uncertainty $38.1\pm1.0$\,s.

In addition to the seismic mean period spacing, our method requires an absolute magnitude determined from astrometry.
Pulsating white dwarfs that are bright enough for precision high-speed photometry from \textit{Kepler}, \textit{TESS}, or the ground, are nearby enough to have precisely measured \textit{Gaia} parallaxes.  The parallax value measured for WD\,0158$-$160 (\textit{Gaia} DR3 \verb+source_id+ 5147930591051748480), for instance, is $\varpi = 14.5715\pm0.0373$\,mas. The re-normalised unit weight error (RUWE) value for this target is 1.049, implying that the single-star model is a good fit to the astrometry \citep{El-Badry2024}. We subtract a zero-point correction of -0.02933\,mas based on \citet[as calculated by \citealt{GentileFusillo2021}]{Lindegren2021}. For a nearby star with small fractional error on parallax, the simple distance estimate based on the Taylor expansion of $d = 1000/\varpi \pm 1000\times\sigma_\varpi/{\varpi^2}$\,pc is valid. This yields $d = 68.49\pm0.18$\,pc for WD\,0158$-$160, in close agreement with the estimate from \citet{Bailer-Jones2021} report $68.52^{+0.19}_{-0.20}$\,pc that uses a smooth distance prior based on a model of the Galaxy.

\textit{Gaia} DR3 cataloged a very precise mean \textit{G}-band magnitude for WD\,0158$-$160 of $14.6567\pm0.0011$\,mag. 
The distance modulus is applied to obtain the extincted absolute $G$-band magnitude: $G - 5\log_{10}{d} + 5 = 10.479\pm0.006$\,mag, with errors propagated from both distance and apparent magnitude.
To obtain an absolute magnitude that is comparable with stellar models, we must apply an extinction correction; the effect of extinction to cause a star to appear fainter is degenerate with the star being cooler or smaller. While extinction is often treated as negligible within 100\,pc \citep[e.g.,][]{Harris2006,Genest-Beaulieu2019}, the 3D extinction maps of \citet{Vergely2022} predict an extinction correction for \targ\ of $A_V = 0.010$\,mag with negligible reported uncertainty, as calculated by \citet[details therein]{GentileFusillo2021}. Prior to the availability of the \citeauthor{Vergely2022} maps, \citet{GentileFusillo2019} estimated an extinction for this target of 0.014\,mag based on the dust maps and corrections of \citep{Schlegel1998} and \citep{Schlafly2011}. We adopt the difference between extinction estimates of 0.004\,mag as the uncertainty on the extinction correction, $A_V = 0.010\pm0.004$\,mag; therefore, the extinction is consistent at the 3$\sigma$ level with both 0.0 and the median extinction of 0.020\,mag estimated for all white dwarfs within 100\,pc with the \citeauthor{Vergely2022} maps by \citet{GentileFusillo2021}. In the following analysis, we use the value $M_G = 10.469\pm0.007$\,mag for the absolute $G$-band magnitude of \targ .

\section{The Models}\label{sec:models}

The seismic technique developed in this work compares the observables of mean period spacing (\deltaP) and absolute $G$-band magnitude (\gmag) to values calculated for a grid of stellar models. These observables are the focus of the seismic technique because they vary monotonically across the parameter space of mass (\mstar) and effective temperature (\teff), so that values can be reliably interpolated between points on the model grid. We use the White Dwarf Evolution Code \citep[WDEC;][]{Bischoff-Kim2018} to generate structural models that span the parameter space of the DBV instability strip relevant to the analysis of \targ. We sample \mstar\ in the range 0.45--0.90\,\msun\ in steps of 0.05\,\msun, and \teff\ from 20,000--30,000\,K in steps of 500\,K.

\begin{figure*}
	\centering
	\includegraphics[width=0.99\columnwidth]{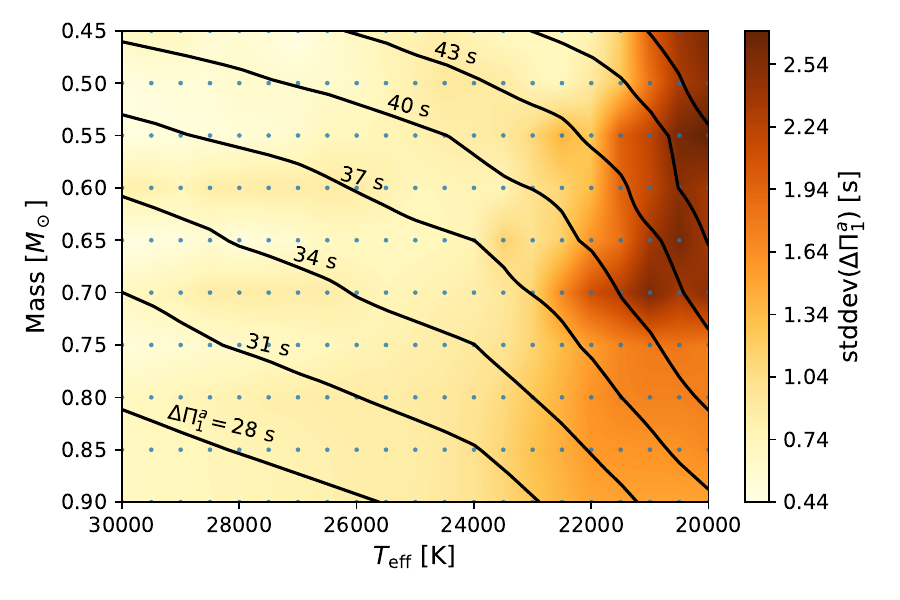}
    \includegraphics[width=0.99\columnwidth]{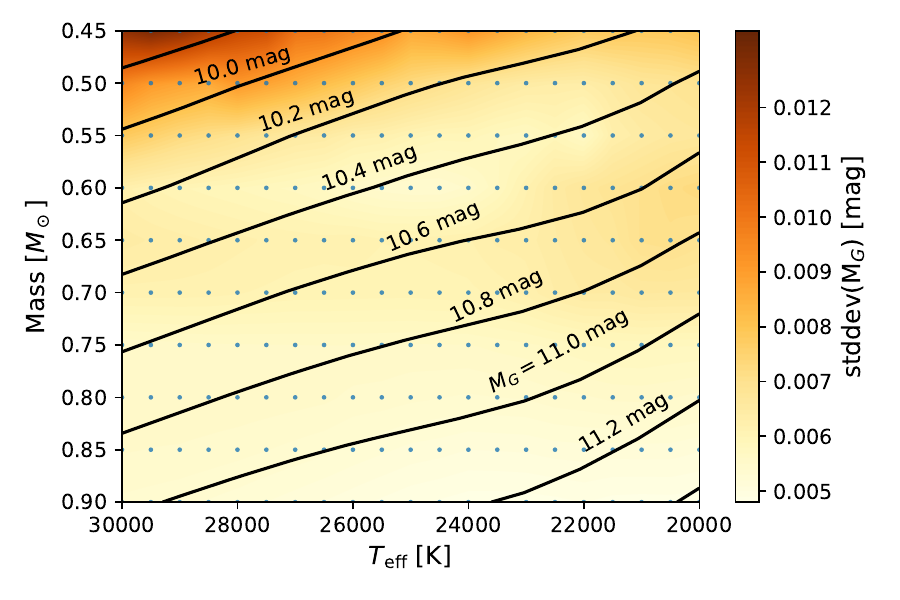}
	\caption{Contours indicating how the distributions of mean period spacings of $\ell=1$ modes (left) and absolute $G$-band magnitudes (right) vary across the mass--effective temperature space of our DBV model grid. The grid of dots indicates the combinations of mass and \teff\ where we computed a set of stellar models with a range of plausible interior structures. The solid contour lines trace the interpolated medians of the distributions for these observables, and the color contour background represent the standard deviations of the distributions (values given in the color bars). The opposing directions of the contours of constant period spacing and absolute magnitude are synergistic for constraining global stellar parameters of mass and \teff.}
	\label{fig:models}
\end{figure*}

To propagate uncertainty from the unknown interior structure of \targ\ to our final results, we consider models with a range of interior structures. WDEC generates structural models with user-specified parameterized core and envelope compositions \citep{Bischoff-Kim2018}. While the global parameters of \mstar\ and \teff\ are the primary influences on \deltaP\ and \gmag, the interior composition does have some effect on the observables. At each grid point of \mstar\ and \teff, we compute a set of  models with interior structures that span a range of configurations that are physically plausible based on the results of fully evolutionary stellar model calculations. The details of the 320 interior structures that we aim to compute for each point in our grid are provided in Appendix~\ref{app:models}. 
By sampling parameters at the extremes of what is likely to be relevant to an actual DBV pulsator, we obtain histograms of observables \deltaP\ and \gmag\ from which we can measure medians and standard deviations. These characterize the distributions of observables that are compatible with models of different masses and effective temperatures for our comparison statistics.

The mean period spacings are calculated for each model as the slope of the best-fit line to the $\ell=1$ modes versus radial order. The periods are expected to be distributed about an even spacing in period on average in the asymptotic limit where $k \gg \ell$ \citep{Tassoul1990}. We used WDEC to compute pulsation periods for modes from $k = 1$ to modes with periods up to 1500\,s. Observed periods of average-mass pulsating white dwarfs rarely exceed the 1500-s upper limit. Based on visual inspection of plots of period versus $k$, we determined that modes with $k \geq 10$ are safely in the asymptotic limit and not severely affected by mode trapping \citep[in agreement with the findings of ][]{Corsico2002}, and we fit the line to only those modes.
To confirm the expected linear behavior of the period differences, we apply a validation test that the average root sum of squared residuals per period is less than a fixed empirical threshold for $\ell=1$ modes of 6.0\,s.
This limit was determined by visual inspection of the best-fit lines through the data points. If the normalized scatter exceeds this limit or fails to be strictly positive, the fit is flagged as poor, the data is rejected as non-linear or anomalous, and the result is discarded. 99.7\% of the models in our grid pass this test.

To compute absolute $G$ band magnitudes for the WDEC models, we reference synthetic magnitudes computed for the Gaia DR3 $G$ bandpass from white dwarf atmosphere models available online.\footnote{\url{https://www.astro.umontreal.ca/~bergeron/CoolingModels/}}
The magnitudes are computed following the methodology of \citet{Holberg2006} for the DB model atmospheres of \citet{Bergeron2011}, with masses determined based on the evolutionary sequences of \citet{Bedard2020}. The structures of the models used to compute the synthetic magnitudes will generally differ from the variety of model structures we computed with WDEC, resulting in slight differences in model radii. The absolute magnitude is sensitive to stellar radius, $R_\star$, as the stellar luminosity scales as $L_\star\propto R_\star^2T_{\rm eff}^4$. It is the subtle effect of interior structure on radius (at the 0.3\% level) that produces a distribution of model absolute magnitudes at each point of \mstar\ and \teff\ in our grid. We interpolate the synthetic photometry models for values of $G$-band absolute magnitude, $M_{G,\text{Bergeron}}$, and corresponding reference radius, $R_\text{Bergeron}$, at the effective temperature and surface gravity of every model in our grid. The model magnitude $M_{G,\text{Bergeron}}$ represents the emergent spectrum expected from these models, but we must scale the absolute magnitude to account for the radius of each WDEC model, $R_{\text{WDEC}}$, as
\begin{equation}
M_{G} = M_{G,\text{Bergeron}} -2.5 \log \left( \frac{R_{\text{WDEC}}}{R_\text{Bergeron}}\right)^{2}.
\label{eq:interpolation}
\end{equation}

Not all of the 320 WDEC models that we attempted to compute for each point in our \mstar,\teff\ grid converged or produced the expected profiles. For the most successful combinations of \mstar\ and \teff, there are at most 315 models with different interior structures that converged. In the worst cases, there are as few as 278 converged models. The median number of models per grid point is 309. There are systematics in where models converge, as we have fewer successful models for less massive white dwarfs. In particular, the less massive models that fail are those with deeper, pure-helium envelopes. To assess the impact of these failed models on our distributions of observables, we compare the medians and standard deviations of the most successful grid points at high mass to the results if we drop the combinations of interior structures that fail at lower mass. Omitting these models from the distributions has a practically negligible effect on standard deviation ($<10\%$), but small systematic effects on the medians: the median mean period spacing increases by less than 0.1\,s on average, and the absolute magnitudes decrease by $\lesssim0.0004$\,mag. These are universally smaller than the associated errors from the standard deviation of model observables and should have minimal impact on our final results. In addition, fewer than 3\% of the models that converged showed anomalous mean period spacing values that resulted from nonphysical chemical profiles; these models are excluded from our analysis by clipping all period spacing values more than 4$\sigma$ from the median when measuring the standard deviations. 

We interpolate medians and standard deviations onto a finer grid to resolve solutions. Essentially, we only need the models to sample mass and temperature finely enough to resolve important features in the dependence of the observables. The monotonic trends of these observables across \mstar\ and \teff\ make them straightforward to interpolate onto an arbitrarily fine grid. We interpolate with step sizes that cause changes in observables that are smaller than our measurement uncertainties. Contour plots indicating the response of \deltaP\ and \gmag\ distributions to changes in \mstar\ and \teff\ are shown in Figure~\ref{fig:models}. The solid contour lines represent the medians of the distributions, and the color contours in the background represent the standard deviations.  The fact that the contours of constant \deltaP\ and \gmag\ vary across the parameter space of \mstar\ and \teff\ in opposite directions demonstrates their synergy for uniquely constraining these properties of pulsating white dwarf stars.

\section{Analysis}\label{sec:methods}

\begin{figure}
	\centering
	\includegraphics[width=0.95\columnwidth]{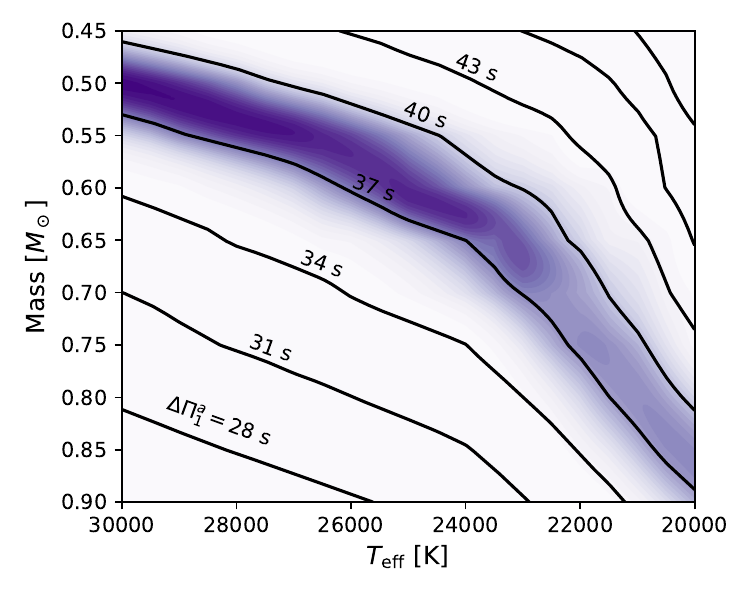}
    \includegraphics[width=0.95\columnwidth]{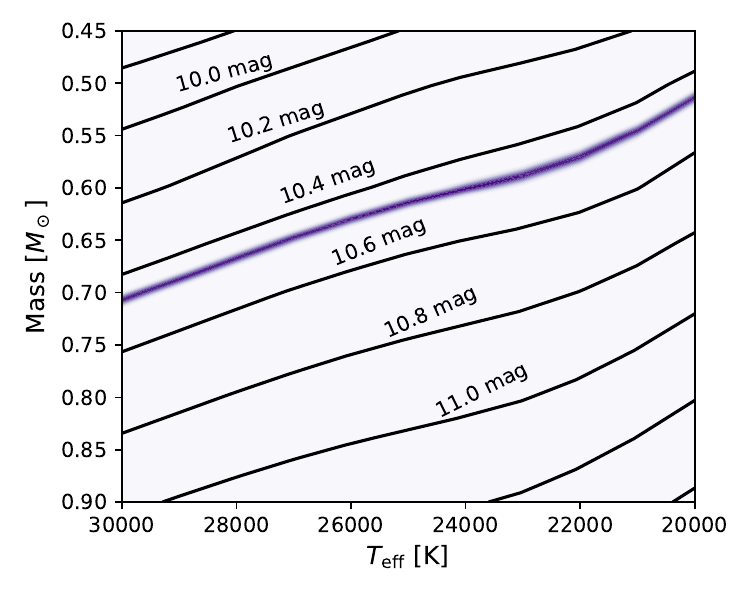}
    \includegraphics[width=0.95\columnwidth]{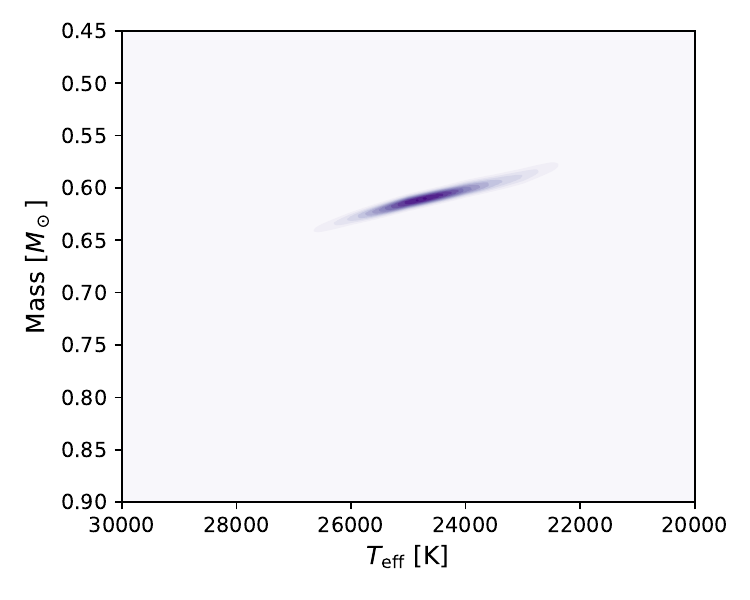}
	\caption{Purple contours show the relative likelihood distribution across mass-\teff\ space for the DBV pulsator \targ\ constrained by the measured mean period spacing (top; \deltaP\ $= 38.1\pm1.0$\,s), absolute $G$-band magnitude (middle; $M_G = 10.469\pm0.006$\,mag), and both measurements combined (bottom). The solid contour lines trace the medians of distributions for these observables interpolated across our model grid from Figure~\ref{fig:models}. }
	\label{fig:likelihoods}
\end{figure}

We compare the distributions of median \deltaP\ and \gmag\ and their associated standard deviations across our model grid to the measurements for \targ\ 
(\deltaP\ $= 38.1\pm1.0$\,s, $M_G = 10.469\pm0.006$\,mag)
to constrain its mass and effective temperature. At each (\mstar,\teff) point in the interpolated model grid, we assess the relative likelihood of the measurements being compatible with the model distributions as
\begin{equation}\label{eq:likelihood}
\begin{aligned}
    L_{ij} \propto
&\frac{1}{\sqrt{\sigma_{\Delta \Pi,ij}^2+\sigma_{\Delta \Pi, \rm obs}^2}}
\exp\left[
-\frac{(\Delta \Pi_{\rm obs}-\overline{\Delta \Pi}_{ij})^2}{2(\sigma_{\Delta \Pi,ij}^2+\sigma_{\Delta \Pi, \rm obs}^2)}
\right]  \\
 \times&\frac{1}{\sqrt{\sigma_{M_G,ij}^2+\sigma_{M_G,\rm obs}^2}}
\exp\left[
-\frac{(M_{G,\rm obs}-\overline{M_G}_{ij})^2}{2(\sigma_{M_G,ij}^2+\sigma_{M_G,\rm obs}^2)}
\right],
\end{aligned}
\end{equation}
where $i$ and $j$ represent one point in the (\mstar,\teff) grid, $\overline{\Delta \Pi}_{ij}$ and $\overline{M_G}_{ij}$ are median values of the distributions at each grid point with associated standard deviations $\sigma_{\Delta \Pi,ij}$ and $\sigma_{M_G,ij}$, and $\Delta \Pi_{\rm obs}$ and $M_{G,\rm obs}$ are the measurements for \targ\ with associated uncertainties $\sigma_{\Delta \Pi,\rm obs}$ and $\sigma_{M_G,\rm obs}$. We assume Gaussian statistics in this formulation. We treat the likelihood distribution as a probability distribution for the parameters \mstar\ and \teff\ by normalizing it as
\begin{equation}\label{eq:normalize}
P(M_{\star,i},T_{{\rm eff},j}) = \frac{L_{ij}}{\sum L_{ij}\Delta M_\star\Delta T_{\rm eff}},
\end{equation}
where $\sum L_{ij}$ is the sum of relative likelihoods across the grid, and $\Delta M_\star$ and $\Delta T_{\rm eff}$ are the spacings in our interpolated model grid. This amounts to an assumption that this pulsating white dwarf star has \mstar\ and \teff\ contained by the grid of models. Our model grid spans most of the DBV instability strip and encompasses the majority of the white dwarf mass distribution \citep{Kepler2019}. The monotonic behavior of our observables across parameter space permits a single solution, and we confirm by inspection that the solution for \targ\ is fully encompassed by our grid.

\begin{figure}
	\centering
	\includegraphics[width=0.99\columnwidth]{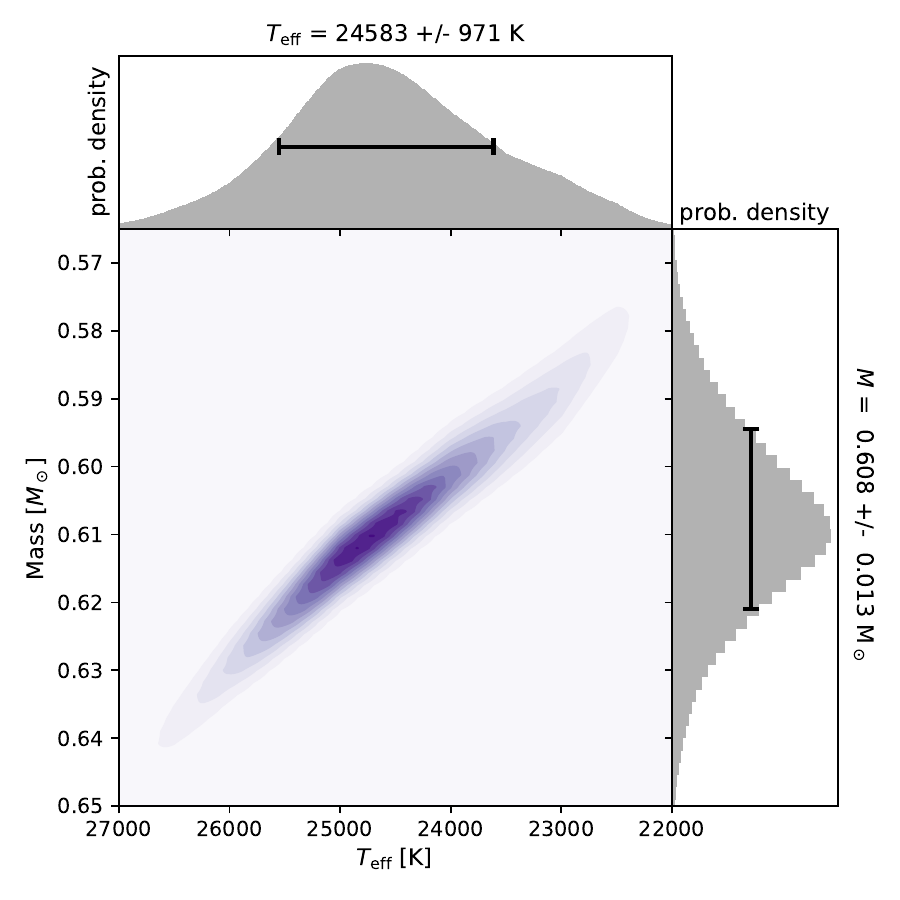}
	\caption{The probability distribution for mass and effective temperature of \targ\ constrained by the seismic technique. The top and right panels display our constraints on each parameter after marginalizing over the other. Error bars mark the 1$\sigma$ range surrounding the median of each marginalized distribution. The unique solution for this DBV star is fully resolved and contained by our grid, and the covariance between parameters is apparent in the angle of the two-dimensional distribution.}
	\label{fig:solution}
\end{figure}

Figure~\ref{fig:likelihoods} displays contour plots depicting the probability distributions for \targ\ constrained through our model grid when considering only \deltaP\ (top panel), when considering only the \gmag\ constraint (middle panel), and their combined constraint (bottom panel). The first two panels consider the two terms in Eq.~\ref{eq:likelihood} separately, and the constraints trace the contours of constant \deltaP\ and \gmag\ that were seen in Figure~\ref{fig:models}. The constraint from mean period spacing appears to have lower amplitude at low temperatures because of the increased sensitivity on interior structure in the models producing broader distributions in this region of the grid (Figure~\ref{fig:models}). The third panel of Figure~\ref{fig:likelihoods} is the full normalized result of Eq.~\ref{eq:normalize}, revealing a single resolved solution for the mass and effective temperature of \targ. Figure~\ref{fig:solution} shows a zoomed-in view of this solution, along with the marginalized distributions. The constraints are strongly correlated, with correlation coefficient 0.972 estimated from the covariance matrix.  We characterize our combined mass and effective temperature constraints as the median of the marginalized distributions with uncertainties taken as half the 68\% interpercentile range. Our final constraints are $T_\mathrm{eff} = 24584\pm 971$\,K and $M_\star = 0.608\pm 0.013$\,\msun.

\section{Discussion and Conclusions}

We have demonstrated a consistent statistical treatment for interpreting seismic mean period spacings alongside astrometric data for the pulsating helium-atmosphere white dwarf \targ. By analogy to the spectroscopic technique \citep{Bergeron1992} that has been used to estimate white dwarf effective temperatures and masses (via surface gravity interpreted through models), we refer to our complementary approach for rich pulsators developed in this work as the seismic technique. We argue that the observables of mean period spacing and absolute magnitude are more reliably interpretable through our models than the precise absorption line profiles that are fitted by the spectroscopic technique, which can be quite sensitive to modeling and observational systematics. However, while the spectroscopic technique can be applied to many white dwarf spectra, the seismic technique is limited to a handful of resolved white dwarfs that not only pulsate, but exhibit a clear sequence of radial overtones. For these stars, parameters obtained with the seismic technique provide valuable benchmarks for testing the performance of spectroscopic and other methods for characterizing white dwarfs.

\begin{figure}
	\centering
	\includegraphics[width=0.99\columnwidth]{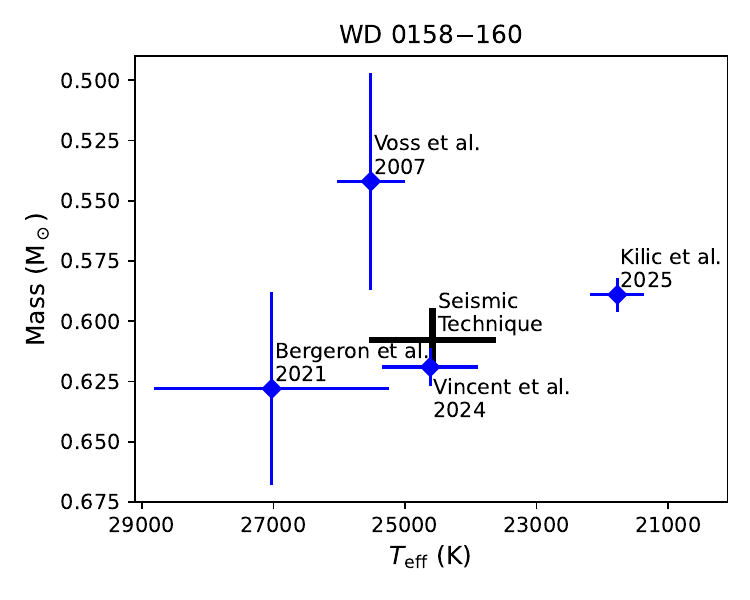}
	\caption{Comparison of mass and effective temperature determinations for \targ\ from ours and other works. See text for discussion.}
	\label{fig:comparison}
\end{figure}

We compare our results to four other sets of mass and temperature measurements for \targ\ that are included on the Montreal White Dwarf Database \citep{Dufour2017}.
\citet{Voss2007} fit DB atmosphere models to spectra obtained by the Supernova Ia Progenitor Survey \citep[SPY;][]{Napiwotzki2003}, obtaining best-fit values of \mstar\ = 0.542\,\msun\ and \teff\ = 25518\,K. They find mean differences between pairs of best-fit parameters for targets that were observed multiple times by the survey (after rejecting outliers) of  $\Delta T_{\rm eff}/T_{\rm eff} = 2.03\%$ and $\Delta \log{g} = 0.058$\,dex \citep{Voss2007}, which we plot as the typical 1$\sigma$ errors. Assuming a simple analytical mass-radius relationship for typical white dwarfs of $M_\star\propto R_\star^{-3}$, this \logg\ error corresponds to a mass error or $\Delta M_\star/M_\star = 8.3\%$. \citet{Bergeron2021} apply the spectroscopic technique to spectra from the Montreal-Cambridge-Tololo colorimetric survey \citep{Demers1986}, obtaining \teff\ $= 27,020\pm1788$\,K and \mstar\ = $0.63\pm0.04$\,\msun. 
\citet{Kilic2025} apply the photometric technique \citep{Bergeron2019}, fitting models to the spectral energy distribution of archival survey photometry with astrometric constraints from \textit{Gaia}, obtaining \teff~$=	21773\pm410$\,K and \mstar\ $=	0.589\pm0.007$\,\msun. 
\citet{Vincent2024} integrate low-resolution \textit{Gaia} XP spectra over photometric filter curves and apply the photometric technique to obtain \teff\ $=24617\pm731$\,K and \mstar\ $= 0.619\pm0.008$\,\msun. Figure~\ref{fig:comparison} compares the results of our work to these other estimates of mass and effective temperature for \targ.  Our results are in excellent agreement with the photometric constraints of \citeauthor{Vincent2024}, and within 3$\sigma$ of the \citeauthor{Kilic2025} result. Application of the seismic technique to a larger sample of stars will allow for the characterization of any systematics between methods \citep[as in][]{Calcaferro2024}.

The photometric technique results of \citet{Vincent2024} and \citet{Kilic2025} are not independent of our findings, as they also interpret the \textit{Gaia} parallax and photometry. Those three results are aligned near a contour of constant absolute magnitude that is strongly constrained by parallax. 
These works utilize the wavelength dependence of different spectroscopic or photometric data sets to essentially position the star along the astrometry-constrained track through \mstar--\teff\ space, while our technique uses asteroseismology. 
These works take different approaches to estimating extinction corrections that will have a systematic effect on the results (e.g., \citeauthor{Kilic2025}\ assume no extinction for targets within 100\,pc). Likewise, we treat the synthetic model magnitudes \citep{Holberg2006} as accurate, as is typical of the photometric technique. In applying the photometric technique to multi-band SDSS photometry, \citet{Genest-Beaulieu2014} found excess residuals over the reported measurement errors of $\approx0.02$\,mag; while they attribute this to excess noise in the measurements, it likely includes some systematic modeling error in the synthetic photometry.  Our results will inherit any systematic error in modeled emergent white dwarf flux or from the calibration of the \textit{Gaia} passband \citep{Riello2021}.

A crucial idea behind our seismic technique is the propagation of uncertainties from the unknown interior structure of the white dwarf. We generate distributions of observable \deltaP\ and \gmag\ values that could arise from plausible interior structures. Parameterized structural models from codes like WDEC are capable of producing physically unlikely configurations; we restrict our exploration of interior structures that span a range of profiles produced by full evolutionary calculations \citep[especially those that explore different plausible implementations of physics, e.g.,][]{DeGeronimo2017}. Our scheme for sampling interior structures relevant to DBV stars for this work is detailed in Appendix~\ref{app:models}.

Relating a readily measurable seismic metric like mean period spacing to global stellar parameters is one of the most robust and widely applied approaches in asteroseismology. The period-luminosity relationship for Cepheid variables established their utility as standard candles \citep{Leavitt1908,Fernie1969}, for example.
Ratios of the dominant periods in high-amplitude pulsators place them on Petersen diagrams \citep{Petersen1973}, and can be used to further constrain metallicity of Cepheids \citep{Lemasle2018} and high-amplitude $\delta$ Scuti variables \citep{Petersen1996}. 
For solar-like oscillations stochastically driven by turbulence in an outer convection zone, scaling relations allow masses and radii to be estimated directly from measurements of the frequency of maximum oscillation power and the large frequency separation between adjacent pressure-mode overtones \citep{Kjeldsen1995,Hekker2020}. The seismic technique presents a similar opportunity to pin down global stellar parameters for white dwarf stars from the mean period spacing, which is the gravity-mode analog to the large frequency separation. 

Having demonstrated the methodology of our seismic technique, we plan to apply it to roughly a dozen richly pulsating white dwarfs exhibiting pulsation periods consistent with a radial overtone sequence \citep[e.g.,][]{Bischoff-kim2024} in future work. 
Appropriate targets are those exhibiting multiple long-period pulsations that align closely with a sequence with near-constant period spacing, characteristic of sequential radial orders in the asymptotic regime ($k\gtrsim 10$) potential bias from the effects of mode trapping on the mean period spacing measurements is minimal \citep{Corsico2002}. 
We will extend our modeling effort to cover the ZZ Ceti instability strip containing the hydrogen-atmosphere pulsating white dwarfs (DAVs). For DAVs, the hydrogen layer mass has a significant impact on the radius \citep{Romero2019} and period spectrum \citep{Bischoff-Kim2023,Uzundag2023} and will need to be sampled.

A further goal of white dwarf asteroseismology is to study white dwarf interior structures by fitting individual measured frequencies to stellar models \citep[e.g.,][]{Giammichele2022}. Doing so with parameterized models like those from WDEC is computationally expensive, as it requires resolving solutions in a high-dimensional parameter space. 
Using the seismic technique to considerably narrow the region of \mstar--\teff\ space consistent with astrometry and a measured mean period spacing can greatly increase efficiency by informing what models to compute. The seismic technique may help realize the promise of white dwarf asteroseismology to reliably map white dwarf interior structures.

\begin{acknowledgments}
This material is based upon work supported by the National Science Foundation under Award AST-2406917.
 This work has made use of data from the European Space Agency (ESA) mission
{\it Gaia} (\url{https://www.cosmos.esa.int/gaia}), processed by the {\it Gaia}
Data Processing and Analysis Consortium (DPAC,
\url{https://www.cosmos.esa.int/web/gaia/dpac/consortium}). Funding for the DPAC
has been provided by national institutions, in particular the institutions
participating in the {\it Gaia} Multilateral Agreement.
This paper includes data collected by the TESS mission. Funding for the TESS mission is
provided by the NASA Explorer Program. This research has made use of the SIMBAD database,
operated at CDS, Strasbourg, France. This work made use of the Montreal White Dwarf Database \citep{Dufour2017}.
We thank the anonymous referee for constructive feedback on this manuscript.
\end{acknowledgments}

\appendix
\section{Interior structure parameters for the model grid}\label{app:models}

We construct a grid of WDEC \citep[version 20]{Bischoff-Kim2018} models spanning the DBV instability strip with a variety of interior structures. The goal is to sample the distributions of mean period spacings and absolute $G$-band magnitudes that we could reasonably expect to exist in nature, informed by the results of stellar evolutionary models. We only need enough models to robustly estimate the medians and standard deviations of these distributions for use in our likelihood statistics (Eq.~\ref{eq:likelihood}). 

WDEC generates structural models with interior profiles parameterized to specify the core oxygen composition profile and the helium profile (details in \citealt{Bischoff-Kim2018} and \citealt{Bischoff-Kim2018b}). There is also one parameter tied to any hydrogen in the model, the mass of the hydrogen layer; for DBVs, that last parameter is set to a numerical zero. Instead of considering single interior structures that one might expect from evolutionary calculations, we sample parameters sparsely through physically plausible ranges of combinations to convolve their effects together into distributions of observables. We characterize those distributions with measured medians and standard deviations. Varying some of these parameters has a noticeable effect on the mean period spacing of $\ell=1$ modes or on the absolute magnitude (by affecting the radius), while others are practically negligible. Constructing a distribution of observables from models with a range of interior structures allows us to essentially marginalize over these effects for reasonable error propagation. 

To determine which parameters have a significant effect on the observables, we sampled each through a range that appears achievable for DBVs through fully evolutionary calculations \citep[e.g.,][]{Althaus2010,DeGeronimo2019}. For each varied parameter, we perform Kolmogorov-Smirnov (K-S) tests on the distributions of model mean period spacing and radii that result from different values of the parameter to determine which are most important. We choose to sample in our model grid all parameters that cause a K-S statistic anywhere across our grid to exceed 0.3 in our initial tests of parameter sensitivity. At a given mass and effective temperature, the most impactful parameters are the mixing-length theory (MLT) $\alpha$ and pure helium layer depth ($M_{\rm He}$; more helium results in a larger atmosphere). $\alpha$ affects the depth of the convection zone, and that is the outer turning point for the longer period modes. Other parameters that set the shape of the helium envelope also have a notable impact:  $M_{\rm env}$ (the location of the base the helium envelope, where the helium abundance first begins to become non-zero when moving outward from the core), as does $\alpha_1$ (a diffusion parameter, though $\alpha_2$ is negligible) and $X_{\rm He}$ (the helium abundance in the mixed C/He region). For the core oxygen profile, the abundances in different regions are important in the context of this work ($h_1, h_2,h_3$) but not the locations of their transitions ($w_1, w_2,w_3,w_4$).

For mixing length parameter $\alpha$ \citep[ML2; ][]{Bohm1971}, we consider the range of values advanced by various studies as being relevant to modeling the convection zones of DBVs. \citet{Bischoff-Kim2015} explored the effect of the mixing length parameter on the fitting of DBVs, finding that $\alpha$ affects the modes particularly for cool pulsators that tend to exhibit longer-period modes that are bounded by the base of the convection zone. \citet{Provencal2015} report results for MLT $\alpha$ between 0.67--1.24 derived from fitting DBV light curves with models of nonlinear interactions between pulsations and the convection zone \citep[see Table 2 of][]{Bischoff-Kim2018}. \citet{Cukanovaite2019} calibrate values of mixing length $\alpha$ that match results of 3D hydrodynamic simulations of convection in DB white dwarfs \citep{Cukanovaite2018}. Values of $\alpha$ in the range 0.4-1.2 appear relevant to the DBV instability strip with effective temperatures in the range $\approx20{,}000-30{,}000$\,K. We sample WDEC models with values of $\alpha$ = [0.4,0.6,0.8,1.0,1.2] at each grid point to not overconstrain our results by assuming a specific convective efficiency.

In WDEC, the envelope profile is parameterized in terms of the mass coordinate $q(M_r) = \log_{10}(1 - M_r/M_\star)$, where $M_r$ is the mass contained within radius $r$. 
This mass coordinate is chosen because helium and hydrogen layers on white dwarfs are very thin (constituting the outer $\sim 1 \%$ of the model by mass) and $q$ conveys structure in the envelope better. In what follows, $q$ shall denote transitions occurring in the helium envelope. The first one is the base of the envelope itself, $q_{\rm env}$. The second is $q_{\rm He}$, the mass coordinate where the helium abundance rises to 1 (thickness of the pure helium layer). 
Between these points is a plateau with fractional helium abundance $X_{\rm He}$.

\citet{Camisassa2017} calculated evolutionary model sequences for DB white dwarfs, reporting total helium layer masses in their Table~1. 
These total layer masses are comparable with $q_{\rm env}$.  The numbers however, while similar, do not translate exactly, and so in order to reproduce the \citeauthor{Camisassa2017} models, we ran WDEC models with varying values of $q_{\rm env}$ until we obtained a match in helium layer profiles. In \citet{Camisassa2017}, less massive DBs with \mstar~$\approx 0.5$\msun\ have $q_{\rm env} \approx -1.4$, while more massive DBs with \mstar~$\approx 0.9$\msun\ have $q_{\rm env} \approx -3$. The seismic solution of \citet{Giammichele2018} for a parameterized structure model of a 0.570\msun\ DBV had $q_{\rm env} \approx -4$; \citet{DeGeronimo2019} explored the physical plausibility of this result by modifying the physics of evolutionary models, obtaining such small values only for massive DBs of $\approx 1$\msun.

\begin{table}[t]
\centering
\begin{tabular}{||c c c||} 
 \hline
 h1 &  h2 & h3 \\  [0.5ex] 
 \hline\hline
0.650 & 0.325 & 0.228\\
0.650 & 0.325 & 0.276\\
0.650 & 0.585 & 0.410\\
0.650 & 0.585 & 0.49725\\
0.900 & 0.450 & 0.315\\
0.900 & 0.450 & 0.3825\\
0.900 & 0.810 & 0.567\\
0.900 & 0.810 & 0.6885\\ [1ex] 
 \hline
\end{tabular}
\caption{Combinations of parameters for the fractional oxygen abundance at different regions of the core profile as parameterized in WDEC version 20 \citep{Bischoff-Kim2018b}. Models with each parameter combination are computed for each \mstar,\teff\ point in our grid.}
\label{tab:hs}
\end{table}

\citet{DeGeronimo2019} explore what they claim to be a realistic range of diffusion efficiencies, which affects the depth of the pure-He layer mass and the helium abundance in the homogeneous carbon/helium region of the model. 
For very inefficient diffusion, there would be no pure-He layer, and this would not present spectroscopically as a DB. 
For somewhat inefficient diffusion, the pure-He layer would be very thin, perhaps $q_{\rm He} = -7$, and for efficient diffusion the He envelope is essentially pure, with $q_{\rm env} = q_{\rm He}$. 
Based on that work, we consider values of $q_{\rm env}$ = [-1.4, -3.0]. For $q_{\rm env}$ = -1.4, we use both $q_{\rm He}$ = -7.0 and -1.4. For $q_{\rm env}$ = -3.0, we use both $q_{\rm He}$ = -7.0 and -3.0. All of their DBV profiles have values of $X_{\rm He}\approx0.4$ in the C/He transition region, and we fix this parameter to 0.4 in our models. We sample the WDEC diffusion parameter that determines the sharpness of the chemical profile at the base of the helium envelope ($\alpha_1$), using two values of 4 and 16, spanning most of the range found by \citet{Bischoff-Kim2023} to match evolutionary profiles from \citet{Althaus2010}.

For h1 (the central oxygen abundance), \citet{Giammichele2022} evolve models with resolved ``breathing pulses'' that can yield values of h1 as large as 0.92. \citet{DeGeronimo2019} explore how different values of the overshooting parameter used to evolve DB WD cores can result in different central oxygen abundances, finding values for h1 in the approximate range 0.65 to 0.72. To represent our uncertainty on how h1 could affect the observables, sampling two values for h1 at 0.65 and 0.90 seems reasonable.

For h2 and h3, we can sample the ends of the ``Reasonable Ranges'' described in \citet{Bischoff-Kim2023} that produce core profiles consistent with the evolutionary models of \citet{Althaus2010}: h2 at 0.5 and 0.9 and h3 at 0.7 and 0.85. These were quoted for a version of WDEC that defined h2 and h3 and a fraction of h1 and h2, respectively. We convert to the absolute height parameters that WDEC v20 uses, resulting in eight combinations of h1, h2, and h3 given in Table~\ref{tab:hs} that we generate for each combination of other varied WDEC parameters.

The parameters for the radial coordinates of oxygen abundance transitions have negligible effect on the observables compared to the varied parameters, and we fix them to $w_1 = 0.42$, $w_2 = 0.10$, and $w_3 = 45.0$ based on expectations from evolutionary models. Based on the parameters that we do vary, we aim to compute 320 models per \mstar--\teff\ point on our grid.

\bibliography{mybib}{}
\bibliographystyle{aasjournalv7}

\end{document}